\documentclass{iaus}
\usepackage{graphicx}

\title[JD 11.~~Virtual Laboratories and Virtual Worlds] %% short title %%
{Virtual Laboratories and Virtual Worlds}

\author[Piet Hut]   %% give here short author list %%
{Piet Hut}

\affiliation{Institute for Advanced Study, Princeton, NJ 08540, USA
\\ email: {\tt piet@ias.edu}}

\pubyear{2008}
\volume{246}  %% insert here IAU Symposium No.
\pagerange{xxx--xxx}
\jname{Dynamical Evolution of Dense Stellar Systems}
\editors{E. Vesperini, M. Giersz \& A. Sills, eds.}
\begin{document}

\maketitle

\begin{abstract}
Since we cannot put stars in a laboratory, astrophysicists
had to wait till the invention of computers before becoming
laboratory scientists.  For half a century now, we have been
conducting experiments in our virtual laboratories.  However,
we ourselves have remained behind the keyboard, with the screen
of the monitor separating us from the world we are simulating.
Recently, 3D on-line technology, developed first for games but
now deployed in virtual worlds like Second Life, is beginning
to make it possible for astrophysicists to enter their virtual
labs themselves, in virtual form as avatars.  This has several
advantages, from new possibilities to explore the results of
the simulations to a shared presence in a virtual lab with
remote collaborators on different continents.  I will report
my experiences with the use of Qwaq Forums, a virtual world
developed by a new company (see http://www.qwaq.com).
\end{abstract}

\firstsection % if your document starts with a section,
\section{Introduction}

A year ago, I was vaguely familiar with the notion of virtual worlds.
I had read some newspaper articles about Second
Life,\footnote{http://secondlife.com} which seemed
mildly interesting, but I had no clear idea about what it would be
like to enter such a world.  All that changed when I was invited to
give a popular talk on astronomy, in
Videoranch\footnote{http://www.videoranch.com},
another much smaller virtual world.  I realized how different this
type of medium of communication is from anything I had tried before,
whether telephone or email or instant messaging or shared screens.
There was a sense of presence together with others that was far more
powerful and engaging than I had expected.  I quickly realized the
great potential of these worlds for remote collaboration on research
projects.

Since then, I have explored several virtual environments, with the
aim of using them as collaboration tools in astrophysics as well as
in some interdisciplinary projects in which I play a leading role.
By and large my experiences have been encouraging, and I expect
these virtual worlds to become the medium of choice for remote
collaboration in due time, eventually removing the need for most,
but not all, long-distance travel.  The main question seems to be
not so much whether, but rather when this will happen.  My tentative
guess would be five to ten years from now, but I may be wrong: the
technology is evolving rapidly, and things may change even sooner.

In any case, I predict that ten years from now we will wonder how
life was before our use of virtual worlds, just like we are now
wondering about life before the world wide web, and the way we
were wondering ten years ago about life before email.

\section{What is a Virtual World}

Twenty years ago, there was a lot of hype about virtual reality,
with demonstrations of people wearing goggles for three-dimensional
vision and gloves that gave a sense of touch.  These applications have
been slow to find their way into the main stream, partly because of
technical difficulties, partly because it is neither convenient
nor cheap to have to wear all that extra gear.

In contrast, a very different form of virtual reality has rapidly
attracted millions of people: game-based technology developed for
ordinary computers, without any need for special equipment.  About
ten years ago, on-line 3D games, shared by many users, made their
debut.  In such a game, each player is represented by a simple
animated figure,
called an avatar, that the player can move around through the
three-dimensional world.  What appears on the screen is a view of the
virtual world as seen through the eyes of your avatar, or as seen from
a point of view several feet behind and a bit above your avatar, as
you prefer.

In this way, a virtual world is a form of interactive animation movie,
in which each participant plays one of the characters.  Currently, many
millions of players take part in these games, the most popular of
which is World of Warcraft\footnote{http://www.worldofwarcraft.com}.
In addition, other virtual worlds have sprouted up that have nothing
to do with games, or with killing dragons or other characters.
Players enter these worlds for social reasons, to meet people to
communicate with, or to find entertainment of various forms.
Currently the most popular one is Second Life (SL).

A lot has been written about SL, as a quick Google search will show
you.  Businesses have branches in SL, various universities including
Harvard and MIT have taught classes, and political parties in the
French elections earlier this year have been represented there.  SL
has its own currency, the Linden dollar, convertible into real dollars
through a fluctuating exchange rate, as if it were a foreign currency.
In many ways, SL functions like a nation with its own economic, social
and political structure.

\section{Virtual Spaces as Information Tools}

The world wide web has revolutionized global exchange of information.
The notion of global connectivity has been novel, but the arrangement
of content has not proceeded much beyond that of the printed press,
with an element of tv or movies added.  The dominant model is a bunch
of loose-leaved pages, which are connected through a tree of pointers,
allowing the user to travel in an abstract way through the information
structure.  As a result, it is often difficult to retrace your steps,
to remember where you've been, or to take in the whole layout of a site.

In contrast to the abstract nature of the two-dimensional web, virtual
worlds offer a very concrete three-dimensional information structure,
modeled after the real world.  While these worlds are virtual in being
made up out of pixels on a screen, the experience of the users in
navigating through such a world is very concrete.  Virtual worlds call
upon our abilities of perception and locomotion in the same way as the
real world does.  This means that we do not need a manual to interpret
a three-dimensional information structure modeled on the world around
us: our whole nervous system has evolved precisely to interact with
such a three-dimensional environment.

Remembering where you have seen something, storing information in a
particular location, getting an overview of a situation, all those
functions are far more natural in a 3D environment than in an abstract
2D tree of web pages.  One might argue that the technological
evolution of computers, beyond being simply `computing devices', has
moved in this direction from the beginning.  The only reason that it
has taken so long is the large demand on information processing needed
to match our sensory input.

Fortunately, the steady increase in processing power of personal
computers is now beginning to make it possible for everyone to be
embedded in a virtual world, whenever they choose to do so, from the
comfort of their own home or office.  As long as you have a relatively
new computer with a good graphics card and broadband internet access,
there are many virtual worlds waiting for you to explore.  Some of
them, like Second Life, offer a free entry-level membership, only
requiring payment when you upgrade to more advanced levels of
activity.  Getting started only requires you to download the client
program to your own computer; after only several minutes you are then
ready to enter and survey that virtual world.

\section{Virtual Spaces as Collaboration Tools}

Email and telephone have given us the means to collaborate with
colleagues anywhere on earth, in a near-instantaneous way.  Yet both
of them have severe limitations, compared to face-to-face meetings.
In neither medium can you simply point to a graph as an illustration
of a point you want to make, nor can you use a blackboard to scribble
some equations or sketch a diagram.  Three new types of tools have
appeared that attempt to remedy these shortcomings.

One approach is to use video conferencing.  Each person can see one
or more others, in a video window on his or her own computer.  While
this gives more of a sense of immediate contact, compared to a
voice-only teleconference call, it is not easy to use this type of
communication to share any but the simplest types of documents.

Another approach has been to give each participant within an on-line
collaboration access to a window on his or her computer that is shared
between all of them.  Whatever one person types will be visible by all
others, and in many cases everybody is connected through voice as well,
as in a conference call.

The third approach, the use of a virtual world, not only combines some
of the advantages of both, it also adds extra features.  Unlike a
video conference, where participants have rather limited freedom of
movement, virtual worlds offer the possibility of exploring the whole
space.  And in some worlds at least, everybody present in a room can
gather in front of a shared screen that is embedded in the virtual
world, in order to discuss its contents.

\section{Qwaq Forums}

After exploring a few different virtual worlds, I settled on
Qwaq\footnote{http://www.qwaq.com} as
the company of choice for my initial experiment in using virtual
spaces as collaboration tools.  Qwaq is a new start-up company that
provides the user with ready-made virtual offices and other rooms,
called {\sl forums}.
There you can easily put up the contents of various files on wall panels.
Whether they are pdf files, jpeg figures, powerpoint or openoffice
presentations, or even movies,
you can simply drag them with your mouse from your
desktop onto the Qwaq screen and position them on a wall within the
virtual world shown on your screen.  As soon 
as you do that, the file becomes visible for all other users present
in the same virtual room.  The rooms persist between sessions: when
users later visit the same room, your files are still there to be seen.

In addition to such useful files, that can be watched and discussed by
a group of users, Qwaq also allows web browsers to be opened in a wall
panel.  In that way, any piece of information on the web becomes
instantly available for perusing by the participants in a Qwaq forum.
This is not only convenient, it helps give the people present a sense
of embedding and actual presence in the room, given that their whole
discussion takes place in the same virtual space, without a need to
jump out of Qwaq into other applications.  And watching movies
together, avatars can even enjoy meta-virtual presentations within
their virtual environments!

One of the most interesting features is the presence of blackboards
and editors in wall panels.  In this way, users can illustrate their
discussions with drawings and they can type their main points directly
into a file that can be jointly edited by those present.  Later, each
user can easily download a copy of that file onto their own computer.

Almost all of our discussions are held through direct voice communication.
While there is an option for text chatting, the advantage of using a
headset with a microphone to directly talk to each other is so large
that we hardly ever use text.  The main exception is to exchange a few
words while someone is giving a lecture, in order not to interrupt the
speaker, or to ask a question to the speaker which can then be answered
in due time.

The underlying software environment used in Qwaq is Croquet, derived
from Squeak, a language based originally on Smalltalk.  Unlike the
more traditional server-centered virtual world architectures,
Croquet is based on peer-to-peer communication, with potentially
far better scaling properties.  Alan Kay, a pioneer of the 2-D
windowing system for personal computers, was the primary visionary
behind the Croquet system, which now has accrued a thriving community
of open source contributors.

\section{Two Experiments}

After I learned about Qwaq, at the
MediaX\footnote{http://mediax.stanford.edu/} conference at Stanford in
April 2007, I started two independent experiments by launching two
independent initiatives, or in Qwaq terminology, two `organizations'.
The first Qwaq organization, later called MICA, was aimed at my
astrophysics colleagues.  The second organization, called WoK Forums,
was aimed at a widely interdisciplinary group of scholars.

\begin{figure}[b]
% \vspace*{-2.0 cm}
\begin{center}
 \includegraphics[width=12cm]{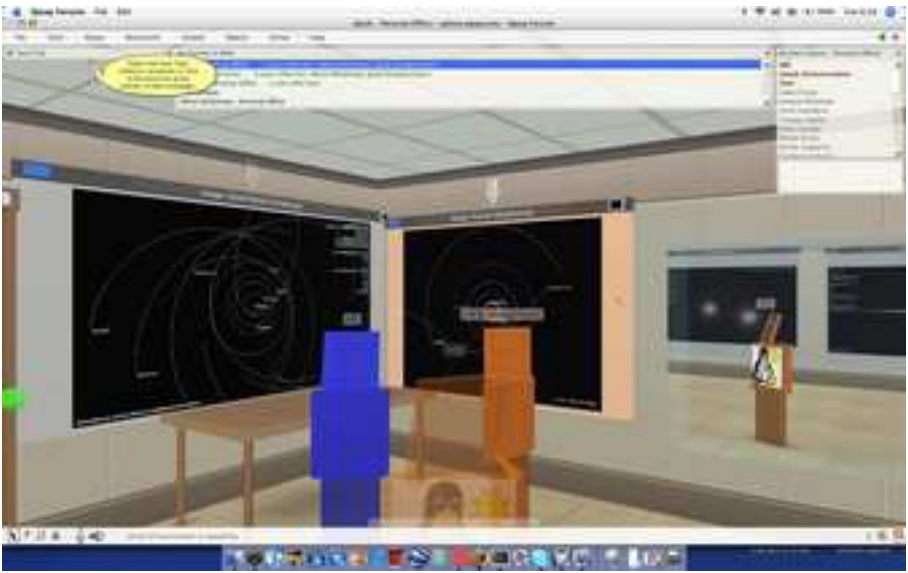} 
% \vspace*{-1.0 cm}
 \caption{An early MICA meeting}
   \label{fig1}
\end{center}
\end{figure}

\subsection{MICA}

MICA stands for Meta-Institute for Computational Astrophysics, with
{\sl meta} derived from the term {\sl metaverse} which is sometimes
used to describe virtual worlds.  During a couple months in the summer
of 2007, we started to explore the use of Qwaq Forums.  One function
was to simply provide a meeting place for people to talk informally, 
a place that can play a role similar to that of a drinking fountain or
a tea room in an academic department.  Other activities were the
organization of seminars and meetings focused on particular topics of
research.

An example of the latter was the MUSE initiative, which stands for 
MUlti-scale MUlti-physics Scientific Environment for simulating dense
stellar systems.  During the
MODEST-7a\footnote{http://www.manybody.org/modest.html}
meeting in Split, Croatia, all
participants of the workshop were given an account in MICA, to give
them a chance to follow up their discussions and collaborations
after the end of the workshop.

\subsection{WoK Forums}

WoK stands for Ways of Knowing\footnote{http://www.waysofknowing.net},
a broadly interdisciplinary
initiative that was started in 2006, with the aim of comparing the
scientific approach to knowledge with other approaches such as those
of art, spirituality, philosophy and every-day life.  For half a year
now, starting in the spring of 2007, we have had daily
meetings in WoK Forums, with many in-depth discussions about notions
such a using your own life as a lab.

\begin{figure}[ht]
% \vspace*{-2.0 cm}
\begin{center}
 \includegraphics[width=12cm]{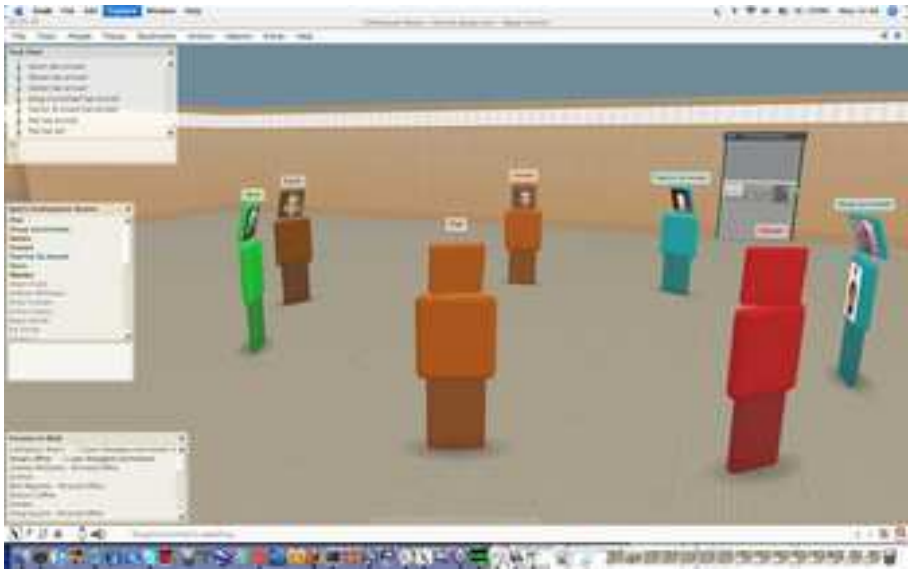} 
% \vspace*{-1.0 cm}
 \caption{A recent WoK Forums meeting}
   \label{fig1}
\end{center}
\end{figure}

Currently we have about two dozen active participants, mostly from
Europe and North America, attending on average one or more meetings a
week.  They range from leading figures in fields such as cognitive
science, psychology, medicine, physics and finance, to graduate
students and postdocs as well as independent scholars and other
professionals.

\section{A Tale of Three Surprises}

When I started the two Qwaq organizations, MICA and WoK Forums,
in May 2007, I did not know what to expect in any detail, given
the novelty of the medium of virtual worlds as a collaborative
tool for academic investigations.  However, I had some rough picture
of what I thought was likely to happen:

\begin{itemize}

\item a quick start for my astronomy group, a slow start for my
interdisciplinary group;

\item virtual worlds as a way to facilitate existing collaborations;

\item an emphasis on using tools: web browsers, 3D objects, etc.

\end{itemize}

\noindent
To my great surprise, all three expectations turned out to be wrong.
What happened instead was:

\begin{itemize}

\item my interdisciplinary group took off right away;

\item I found myself and others creating new collaborations;

\item 3D presence was far more important than specific tools.

\end{itemize}

\subsection{first surprise}

I had expected that the computational astronomers whom I had invited
to MICA would quickly take to the new environment.  After all, most of
them had many years of experience working with rather advanced computer
tools, and many had designed and written their own code and toolboxes.
In contrast, many of the broadly interdisciplinary researchers that I
had gathered were not particularly computer savvy.  I was wondering
whether they would have any interest at all in getting into a new kind
of product that they first would have to download, and then would have
to learn to navigate in.

I was wrong on both counts.  The latter group showed an immediate
interest.  Even though I had started slowly with weekly meetings,
there was strong interest in more frequent gatherings, and soon we
began to meet on a daily basis.  In contrast, the former group, for
whom I had started off with a daily `astro tea time' showed little
interest initially, and most meetings found me being in the tea
room all by myself.

It took a while before it dawned upon me what was happening.  The main
reason was that widely interdisciplinary activities do not have any
traditional infrastructure, in terms of journals, workshops, societies
and other channels to fall back on.  Those people interested in
transcending the borders of their own discipline, not only into the
immediately adjacent discipline but into a range of other
disciplines, have very little to lean on.  By offering a forum for
discussions, I was effectively creating an oasis in a desert,
attracting many thirsty fellow travelers.

In contrast, many of my astrophysics colleagues complain that nowadays
there are already too many meetings and joint activities, and that it
has become increasingly harder to find time to sit down and do one's
own original research, amidst the continuing barrage of email, faxes,
and cell phone conversations.  For them I had created yet one more
fountain in a fountain-filled park.

However, once my astro colleagues started to trickle in, many of them
did find the new venue to be of great interest.  And I had a trick to
increase the trickle: threatening to close MICA sufficed to catch
people's attention, and to increase attendance.  Switching from the
initial daily meetings to weekly meetings also helped considerably.
Having a dozen people in a room discussing the latest news in
computational astrophysics clearly is a lot more fun that being by
yourself or with just one other random person during a daily tea time.

Meanwhile, the daily WoK Forums meetings continue to attract between
half a dozen to a dozen participants on a daily basis, and the
attendance continues to grow.

\subsection{second surprise}

I had expected to kick start my virtual world activities by bringing in
existing teams of collaborators, offering them the chance to continue
what they were doing already, but in a new medium.  Perhaps this new
approach would later attract other individuals, who might be interested
in joining or in starting their own projects, but that was not my initial
objective.

Rarely in my life have I so completely misjudged a situation.
Getting an existing group to make the transition to a totally new mode
of communication turned out to be effectively impossible.  Trying to
change given ways of doing things provoked far more resistance than
I had expected, in both my astrophysics and my interdisciplinary
collaborations.  Simply put, that just didn't work, period.

This became so obvious, very early on, that I had no choice but try a
completely different tag.  I went through my address book, and
gathered names of people who just might have some interest in trying
out a new medium, providing them with some bait, at the off chance
that they might bite.  I had no idea what criterion to use, in order
to attract potential players, given the novelty of the new setup,
so I just threw my net widely, waiting to see what would happen.

Roughly half the people I contacted did not reply.  Of the half that
did reply, roughly half told me that it all sounded fascinating but
that they had no time in the foreseeable future to engage in new fun
and games.  Of the people who did want to give it a try, more than
half quickly got discouraged after trying once or twice, and not
getting immediate gratification one way or another.  But many of those
who remained at the end of this severe selection process were wildly
enthusiastic, considering themselves to be pioneers in a whole new
world.

Even in retrospect, I could never have predicted whom of my colleagues
would fall in the ten percent group of early adopters.  I still do not
see any clear pattern or set of characteristics separating those who
rushed in right away from those choosing to remain sitting on the
fence.  Many of those of whom I had been convinced that they would
embrace virtual worlds did not, and quite a few whom I had contacted
without much expectation turned out to jump in right away.  In fact,
for some of the early players I had not anticipated their interest at
all.  I had contacted them mainly so as not to make them feel left
out when they would hear that I had contacted their seemingly more
promising friends!

Given this randomly hit-or-miss way of collecting early players, any
notion of starting with existing teams rapidly went out the window.
What I wound up with was a bunch of enthusiastic tourists, eager to
look around in the new virtual world that opened up unexpected
horizons, with doing any kind of real work seemingly far from their mind.
They were lured into a new adventure, with new toys.

After a while, though, many of the tourists began to settle down, and
they started to behave more like neighbors.  They began to get to know
each other, although many of them had never met in real life.  Among
the MICA participants, there were some old hands in computational
astrophysics, but there also was a freshly minted PhD in the field of
education, Jakub
Schwarzmeier\footnote{http://home.zcu.cz/\~\,schwarz1/index.html} from
Pilsen, in the Czech Republic, who happened to have written some
astrophysical simulations as part of his educational research.  The
MICA snapshot above shows the room that Jakub created, with me
visiting him together with Alf Whitehead who is a graduate student in
astrophysics in a remote study course in Australia while making a
living as a manager of a team of Ruby programmers in Toronto.  Both
Jakub and Alf had independently contacted me by email, without having
met me in person, less than half a year before I started MICA,
offering their help with my ACS\footnote{http://www.ArtCompSci.org}
project, so it was natural to invite them into MICA.

Finally, some of the tourists that had turned into neighbors finally
began to turn into collaborators.  Seeing each other regularly, and
becoming familiar with each others' interests, they began to spawn new
ideas, some of which led to new projects, with little connection to
the original motivation for them to enter the virtual world where they
had met.  This has happened repeatedly in my interdisciplinary
organization, even though there the discrepancy between people's
background and interests was the largest.  In my astrophysics
organization, the first mile stone was reached when Evghenii Gaburov
and James Lombardi started to write a paper together within MICA,
Evghenii in Amsterdam, Holland, and Jamie in Meadville, Pennsylvania, USA,
which led to a preprint in July 2007 (\cite[Gaburov et al. 2007]
{Gaburov_etal07}).  As far as I know, this is the first astrophysics paper
that has an explicit acknowledgment to a virtual world as the medium
in which it was created.

\subsection{third surprise}

I had expected that the main attraction of a virtual world would
surely be the lure of toys: being able to design and build 3D
objects, to use web browsers in-world, to travel through output of
simulations, all that good stuff.  The Qwaq software designers had
already put an attractive example of a simulation output in their
world, in the form of a simple model of an NaCl crystal.  I had
expected my fellow astronomers to quickly come in with their galaxy
models, following in the footsteps of the Qwaq folks.  

I also had thought they would quickly start playing directly with the
software offered by Qwaq.  In addition to existing applications, Qwaq
offers possibilities for scripting new ones, using Python, and the
underlying Croquet offers even more ways to get into the nuts and
bolts of the whole setup.  I had expected my colleagues, especially
students with more time on their hands, to come in to play like kids
in a candy store.

Once more I was wrong.  In a place full of toys, it was the place
itself, not the collection of toys, that formed a magnet.  The main
attraction for coming into Qwaq Forums was presence.  Presence in a
persistent space, a watering hole that quickly became a familiar
meeting ground, this is what was felt to be the single most important
aspect of the whole enterprise.  Everything else was clearly
secondary.

It goes back to the difference between the abstract nature of the
two-dimensional world wide web, versus the concrete sense of `being
there' that we get when we enter a virtual world.  Hundreds of
millions of years of evolution of our nervous system, in all its
perceptive, motor, and processing aspects, have prepared us for
being at home in a three-dimensional life-like spatial environment.

Sharing such an environment with others turned out to be a factor that
was far more important than I could have guessed.  I, too, was amazed
to experience the difference between a meeting in MICA or WoK Forums,
on the one hand, and being part of a traditional phone conference with
the same number of individuals, on the other.  Teleconference calls
are among the least pleasant chores to be part of, in my work.  It
is not always clear who is talking, there is often little real
engagement, and the whole thing just feels uninspired, leading the
participants to doodling or reading their email or being otherwise
distracted.

In contrast, a meeting of half an hour in a virtual world feels
totally different.  There is a palpable sense of presence.  You can
see where everybody is located, people can move around and gather in
front of a blackboard or poster or powerpoint presentation, and you
can even hear where people are, through the stereo nature of the sound
communication.

\section{Conclusion}

\subsection{Lessons learned}

Of the two groups that I have invited into virtual spaces,
interdisciplinary researchers were the most eager early adopters.
Astrophysicists were much slower to get started, but once they
were in and saw the potential of this new medium, they could
quickly use the infrastructure they already had in their own
field to produce new results, such as writing a preprint within
a virtual space.

Individual early adopters in both groups did not come in as teams.
Instead, they met whoever else was there, behaving first as tourists,
then as neighbors, and only later as potential collaborators,
spontaneously creating new research projects.  In this way,
everything that happened in virtual spaces was serendipitous; trying to get
existing projects moved into virtual spaces encountered too much resistance.
But even these serendipitous activities took place only after significant
encouragement.  To get a group of people to adapt to a new medium seems
to take a considerable and ongoing amount of prodding, using whatever
carrots and sticks one can find.  Trying to organize any type of new
activity in academia resembles the proverbial challenge of `herding cats.'

The main attraction of meeting in a virtual space has turned out to be
the shared presence in a persisting space that the participants sense
and get hooked to.  After a number of meetings with various stimulating
conversations, the regulars want to keep coming back to the familiar
setting, where they know they can meet other interesting people, old
friends as well as new acquaintances.  Being able to visit such a space
at the click of a button is a great asset.  Whether at home or at work,
or briefly logged in at an airport, the virtual space is always there,
and with enough participants, chances are that you will meet people
whenever you log in.  It can function like a tea room in an academic
department, but then in a portable form, always and everywhere within
reach, a curious mix of attributes.

One major obstacle that I have encountered is the fact that the earth
is round.  Never before have I been so conscious of the fact that we
all live in different time zones.  Spatial distances may drop away,
when people meet in virtual spaces, but temporal zone changes don't.
In my interdisciplinary group, where we have experimented for several
months now with daily meetings, I was forced to introduced meetings
twice every day, in order to accommodate the fact that the participants
live on different continents.  In addition to time zone restrictions,
some participants prefer to log in from home in the evening, others
from work during the day.  Scheduling a weekly colloquium has been
rather difficult, with some people forced to get up very early and
others having to stay up till late at night.

As a result of all this, the critical mass needed to sustain a
`tea time' where enough people show up spontaneously is much larger in
a virtual space than it is in an academic department.  With ten people
in a building, and a fixed tea time at 3 pm, chances are that at least
five people show up at any given tea.  With twice-daily meetings in a
virtual space, and many participants showing up only once a week, you
need more like a hundred people in total, to guarantee the presence of
five people per meeting.  And if the attendance often falls below five,
there may not be enough diversity to attract regular attendance.

Trying to organize people to attend events in a virtual space has
something in common both with running a department and with organizing
a workshop.  Like the former, it requires persistent management, unlike
putting together a workshop that is a one-shot event.  And even though
it is much easier to establish a virtual space, compared to getting the
funding and spending the time to build a physical building, it is also
easy to underestimate the time it takes to establish an attractive
infrastructure.  Try to image what it would be like to run a
never-ending workshop, and you get the idea.

In the short run, there is no ideal solution to the management problem.
Trying to run things purely by committee is unlikely to work, nor will
it be easy to find a single individual willing to do the brunt of the
work needed to set up and maintain the infrastructure of a purely virtual
organization for academic research.  Progress is likely to come from
some kind of middle ground, with a small core group of enthusiasts
willing to spend significant amounts of time getting things going, in
a typical `open source' kind of atmosphere, setting the tone by their
personal example.

\subsection{Next steps}

So far, the two organizations that I have founded, MICA and WoK
Forums, are still very much in their initial phase where people are
getting to know each other and are getting to know the virtual
environment and its possibilities.  What will happen next is difficult
to predict.  As always in a new medium, the most interesting
developments will be those that nobody expected.  Even so, there are a
few obvious next steps.

One thing-to-do is to create some form of library or archive,
containing a chronicle of what has happened in a given virtual space.
After people give lectures, it will be good to keep at least their
powerpoint presentations.  When people hold discussions, it would be
great to catch their conclusions in a type of wiki or other structure
for text that is easy to enter.  It would be great if a whole session
in a virtual space could be captured on video, and stored for later
viewing within a room in that same virtual space.

For computational science applications, such as large-scale
simulations in astrophysics, virtual spaces can be at the same time
places for people to meet, and places where those people can run their
experiments.  With individuals represented as avatars, it is natural
for them to enter virtual laboratories where they are running their
simulations.  Instead of the scientist sitting in front of the computer
and the simulation taking place at the other side of the screen, there
are many advantages in letting the scientist enter the screen and the
simulated world directly.  By traveling through a simulation, one can
become much more intimately familiar with the details of a simulation.

Finally, here is one more intriguing possibility.  If researchers who
are geographically remote start writing code together within a virtual
space, we can literally capture all that is said and done while
writing the code.  By keeping the full digital record of a coding
session, and indexing it to the lines of code that were written during
that session, future users of that code will always have the option to
travel back in time to get full disclosure of all that happened during
the writing.  Many of us, struggling with legacy code that was written
decades ago, would be happy to give a minor fortune for the
possibility of making such a trip back in time.  This approach to
massively overwhelming documentation is in the spirit of what
Jun Makino and I have suggested on our Art of Computational Science
website\footnote{http://www.ArtCompSci.org}, as a move from {\sl open
source} to {\sl open knowledge} (\cite[Hut 2007] {Hut07}).

\section*{Acknowledgments}
I thank Sukanya Chakrabarti, Derek Groen, Andrew McGowan, Sean Murphy,
Greg Nuyens, Rod Rees and Patrick St-Amant for their helpful comments
on the manuscript.

\end{document}